\documentclass[aps,prb,twocolumn,groupedaddress,longbibliography,superscriptaddress]{revtex4-1}

\usepackage[utf8]{inputenc}
\usepackage{graphicx}
\usepackage{float}
\usepackage{color}
\usepackage{hyperref}
\usepackage{amsmath}

\begin{document}

%\preprint{}

\title{Ultrafast spin initialization in a gated InSb nanowire quantum dots}



\author{S. Bednarek}
\affiliation{
	Faculty of Physics and Applied Computer Science,
	AGH University of Science and Technology, Krak\'{o}w, Poland}
\author{J. Pawłowski}
\email[]{jaroslaw.pawlowski@pwr.edu.pl}
\affiliation{
	Department of Theoretical  Physics, Faculty of Fundamental Problems of Technology, Wroc\l{}aw University of Science and Technology, Wybrze\.{z}e Wyspia\'{n}skiego 27, 50-370 Wroc\l{}aw, Poland}
\author{M. Górski}
\affiliation{
	Faculty of Physics and Applied Computer Science,
	AGH University of Science and Technology, Krak\'{o}w, Poland}
\author{G. Skowron}
\affiliation{
	Faculty of Physics and Applied Computer Science,
	AGH University of Science and Technology, Krak\'{o}w, Poland}


\date{\today}

\begin{abstract}
We propose a fast and accurate spin initialization method for a single electron trapped in an electrostatic quantum dot. The dot is created in a nanodevice composed of a catalytically grown indium antimonide (InSb) nanowire and nearby gates to which control voltages are applied. Initially we insert a single electron of arbitrary spin into the wire. Operations on spin are performed using the Rashba spin-orbit interaction induced by an electric field. First, a single pulse of voltages applied to lateral gates is used to split the electron wavepacket into two parts with opposite spin orientations. Next, another voltage pulse applied to the remaining gates rotates spins of both parts in opposite directions by $\pi/2$. This way, initially opposite spin parts eventually point in the same direction, along the axis of the quantum wire. We thus set spin in a predefined direction regardless of its initial orientation. This is achieved in time less than $60\,\mathrm{ps}$ without the use of microwaves, photons or external magnetic fields.
\end{abstract}

%\keywords{}

\maketitle

\section{Introduction}
One of the many quantum bit implementations is based on spin of an electron or hole trapped in a semiconductor nanodevice \cite{ref1,ref2,ref3}. Such a device must be built in a way, that allows performing several fundamental operations, namely: initialization, manipulation and readout \cite{ref4,ref5}. Most of them can be easily carried out in electrostatic quantum dots (QDs), for which confinement potentials are generated in quantum wells \cite{ref6,ref7,ref8,ref9,ref10} or wires \cite{ref11,ref12}, by voltages applied to local gates. They are also realised in self-assembled QDs \cite{ref13,ref14,ref15,ref16,ref17}, for which confinement is obtained only due to presence of heterojunctions of different semiconductors. These operations have to be performed sufficiently fast, as a sequence of calculations has to be completed before the decoherence of spin takes place \cite{ref9}.

The most difficult operation to accomplish turns out to be the spin initialization, that is, orienting spin in a chosen direction before any computations are executed. In self-assembled QDs spin of a single electron \cite{ref13,ref14,ref15} or hole \cite{ref16,ref17,ref18} can be set by using optical transitions to excitonic or trionic (charged excitons) states. We can proceed in a similar way in nanowire QDs based on InAsP/InP heterojunctions \cite{ref19}. On the contrary, in electrostatic QDs optical initialization through trionic states is not possible, since an attractive potential for electrons is repulsive for holes and thus, a stable excitonic state cannot be formed. In such systems the Pauli blockade is used \cite{ref9,ref10}. This method allows to set spin of an electron in parallel to spin of another adjacent electron previously trapped in the QD, however spin of the former electron is random. It can be set deterministically by using a strong magnetic field and waiting until the electron relaxes to the ground state \cite{ref10}. Obtained this way initialization is not accurate and takes at least a couple of nanoseconds.
However, to initialize a qubit to a known state for further operations, a high fidelity initialization procedure is necessary \cite{noiri}. Additionally, to perform quantum error correction certain ancillary qubits must be continuously reinitialized in ultra-short (relatively to decoherence) timescales \cite{preskill}.
The main source of electron spin decoherence are interactions with nuclear spin bath. Our nanodevice structure is similar to the one described in the experimental paper \cite{ref12} which invokes a coherence time of about 34 ns.

Recently, in \cite{ref23}, we have designed a nanodevice based on a planar semiconductor nanostructure, which allows for spin initialization with fidelity over 99\%, lasting about $400\,\mathrm{ps}$. This task can be achieved using the electrostatically controlled Rashba spin-orbit interaction (SOI). In this paper we propose a device capable of achieving similar fidelity an order of magnitude faster in a quantum wire, a nanostructure that is well within current experimental capabilities and much easier to fabricate than a planar nanostructure. This makes quantum wires an ideal starting point for experimental reasearch on spin initialization in solid state systems.  

\section{Nanodevice structure}
\begin{figure}[b]
\includegraphics[width=0.4\textwidth]{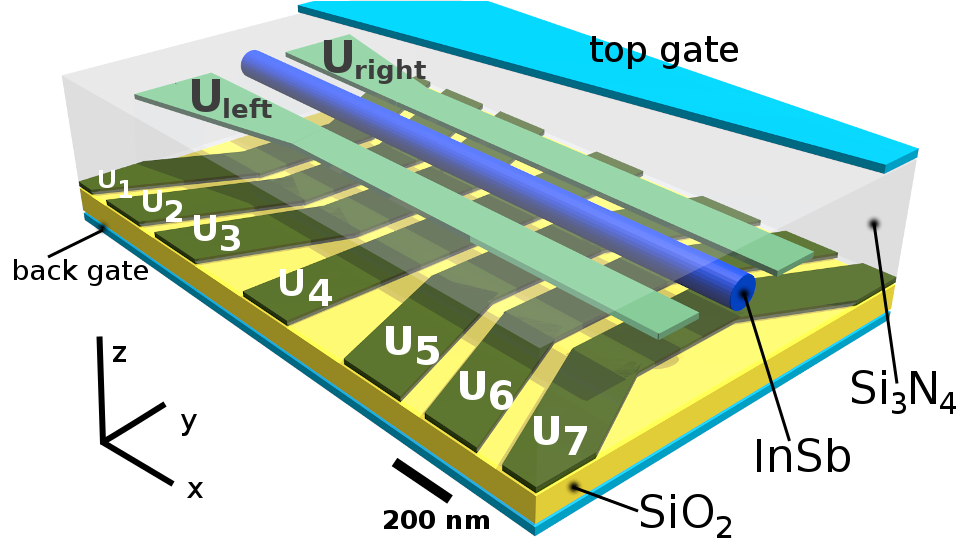}
\caption{\label{fig:nanodevice}Schematic view of the proposed nanodevice containing a gated InSb nanowire. Top gate is shown only partially.}
\end{figure}
For spin initialization we propose a nanodevice similar to those used in \cite{ref24,ref12}, shown in Fig. \ref{fig:nanodevice}. On a strongly doped silicon substrate we place a $100\,\mathrm{nm}$ thick layer of $\mathrm{SiO_2}$. Next, we lay down seven $200\,\mathrm{nm}$ wide metallic gates $\mathrm{U_i}$ separated by gaps of $50\,\mathrm{nm}$ each. They serve to shape the confinement potential along the wire. The gates are then covered with a $260\,\mathrm{nm}$ thick layer of $\mathrm{Si_3N_4}$ insulator. On top of the insulator we put a catalytically grown InSb nanowire, $80\,\mathrm{nm}$ in diameter and of length $L$ exceeding $1.5\,\mathrm{\mu m}$. On both sides of the wire, in parallel, we put two lateral gates $\mathrm{U_{left}}$ and $\mathrm{U_{right}}$ at a distance of $50\,\mathrm{nm}$ from the wire axis. They are used to generate an electric field along the $y$-axis. Everything is then covered with $\mathrm{Si_3N_4}$ up to $400\,\mathrm{nm}$ measured from the substrate. Finally, the top surface of $\mathrm{Si_3N_4}$ is covered with top gate $\mathrm{U_{top}}$, which, along with back gate (formed by the strongly doped substrate) is used to create an electric field parallel to the $z$-direction. 

\section{Model}
The operations on the electron are performed within a range of very low energies near to the conduction band minimum. The initial voltages applied to the gates, and a parabolic approximation of the resulting potential give rise to an excitation energy of about $\hbar\omega=0.2\,\mathrm{meV}$, significantly lower than the InSb band gap of $230\,\mathrm{meV}$. This makes the single band effective mass approximation a reasonable choice. We thus use this approximation in all subsequent calculations. 
Now let us discuss theoretical model of the nanodevice.

First, we assume that the quantum wire confines a single electron. For the InSb effective mass $m=0.014\,m_e$, the energy difference between the ground state and the first excited state of quantized electron motion in perpendicular directions to the wire ($80\,\mathrm{nm}$ in diameter) equals $40\,\mathrm{meV}$, which is two orders of magnitude greater than energies of motion encountered in our nanodevice. Thus, we can use a one-dimensional approximation assuming, that the electron always occupies the ground state of lateral motion.

The corresponding Hamiltonian takes the following form
\begin{equation}\label{eq:hamiltonian}
\mathbf{H}=\left[-\frac{\hbar^2}{2m}\frac{d^2}{dx^2}+V(x)\right]\mathbf{I_2}+\mathbf{H_\mathrm{so}},
\end{equation}
where $V(x)$ is a potential energy of confinement and the last term $\mathbf{H_\mathrm{so}}$ expresses the SOI. The wavefunction takes the spinor form $\mathbf{\Psi}(x,t)=\left(\psi_\uparrow(x,t), \psi_\downarrow(x,t)\right)^\mathrm{T}$. $\mathbf{I_2}$ denotes a $2\times 2$ identity matrix. We assume that the wire is grown along the crystallographic direction $[111]$, thence the Dresselhaus interaction vanishes \cite{ref24, ref25} and we can take into account only the Rashba SOI contribution
\begin{equation}\label{eq:rashba}
\mathbf{H_\mathrm{so}}=\frac{\alpha_\mathrm{so}|e|}{\hbar}\left(E_z\sigma_y-E_y\sigma_z\right)\hat{p}_x,
\end{equation}
with the Pauli matrices $\sigma_y$, $\sigma_z$ and the Rashba coefficient for InSb $\alpha_\mathrm{so}=523\,\mathrm{\AA{}^2}$ \cite{ref25}. $E_y$, $E_z$ are the electric field components within the wire.

If the confinement potential energy has the parabolic form 
\begin{equation}\label{eq:potential}
V(x)=\frac{1}{2}m\omega^2x^2,
\end{equation}
we can solve for the Hamiltonian eigenfunctions analytically in the momentum representation and then transform them to the position representation. Let us assume, that only $E_y$ component of $\mathbf{E}$ is nonzero. The ground state energy is now doubly degenerated with respect to the spin $z$-projection. The wavefunctions in the position representation are gaussians multiplied by plane waves due to the presence of the SOI. Depending on the spin $z$-projection, the wavenumber is either positive $q$ or negative $-q$. Corresponding eigenfunctions take the following form
\begin{equation}\label{eq:eigenstates}
\begin{split}
\mathbf{\Psi}_\uparrow(x)=\frac{2\beta}{\pi}\begin{pmatrix}1\\0\end{pmatrix}e^{-\beta x^2}e^{iqx},\\
\mathbf{\Psi}_\downarrow(x)=\frac{2\beta}{\pi}\begin{pmatrix}0\\1\end{pmatrix}e^{-\beta x^2}e^{-iqx},
\end{split}
\end{equation}
with $\beta=\frac{m\omega}{2\hbar}$ and $q=\frac{m\alpha_\mathrm{so}|e|E_y}{\hbar^2}$. Although each state's wavefunction contains a plane wave factor, the electron remains still as its motion is blocked by the SOI \cite{ref27}. If we now turn off the electric field abruptly by putting $E_y=0$, the SOI vanishes and the electron starts moving according to its momentum to the left ($\langle p_x\rangle=-\hbar q$) or to the right ($\langle p_x\rangle=\hbar q$) depending on its spin. The SOI introduces an energy correction $\Delta E=\frac{\hbar^2q^2}{2m}$, so the maximal electron displacement $\Delta x$ from the equilibrium position of the confinement potential (Eq. \ref{eq:potential}) approximately obeys the relation $\Delta E = V(\Delta x)$, or $\frac{\hbar^2q^2}{2m}=\frac{m\omega^2}{2}\Delta x^2$. It follows that
\begin{equation}\label{eq:displacement}
\Delta x = \frac{\alpha_\mathrm{so}|e|E_y}{\hbar\omega}.
\end{equation}
Converesly, if the electron relaxes to the ground state with $E_y=0$, abrupt turning on of the electric field will set it in motion yet in the opposite direction.
The effect of spin dependent motion when the electric field is turned on can be used for the spin readout. Spin orientation of the electron determines its direction of motion (to the left or to the right). Thus, by measuring the electron presence in the left or in the right half of the wire, after the movement took place, we can infer about its spin orientation. 

If the initial electron state is not an eigenstate of $\sigma_z$ (spin $z$-projection is not definite), its wavefunction is a linear combination of both basis states: $\mathbf{\Psi}(x)=c_\uparrow\mathbf{\Psi}_\uparrow(x)+c_\downarrow\mathbf{\Psi}_\downarrow(x)$. After the electric field $E_y$ is changed, both spinor parts start moving along the $x$-axis in opposite directions and split. If the electric field pulse is sufficiently strong, it is possible to separate them entirely.

\section{Principles of spin initialization}
Before we delve deeper into the details of spin initialization let us first explain the basic principles of this process.
\begin{figure}[t]
\includegraphics[width=0.4\textwidth]{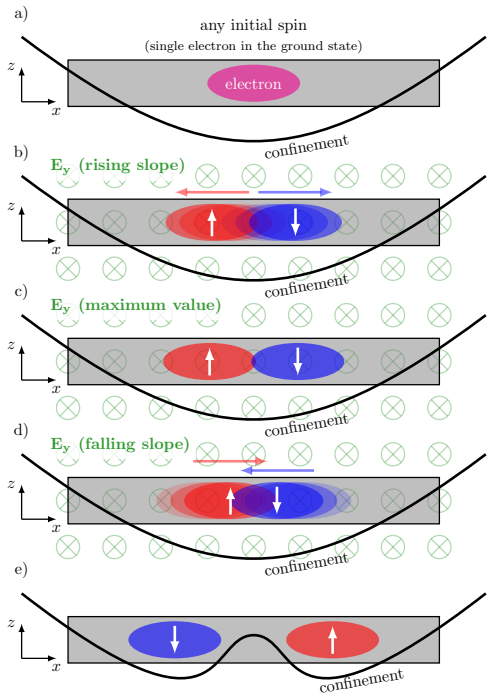}
\caption{\label{fig:stage1principle}First stage of the spin initialization: (a) An electron of arbitrary spin orientation is confined within a quantum wire and occupies the ground state; (b) The rising slope of the electric field pulse $E_y$ along the $y$-axis, due to the Rashba SOI, triggers motion of the electron wavefunction spin components in opposite directions resulting in spin separation; (c) When the spin components are separated the most, the electric field is maximum. The components, however, still overlap and require stronger separation; (d) The falling slope of the electric field pulse $E_y$, employed when the spin components start turning back, accelerates them even more, making them cross each other and reach positions farther apart; (d) Introduction of a potential barrier between the spin components isolates them from their mutual influence and locks their new positions. White arrows along with red or blue colors indicate the spin orientation of each component.}
\end{figure}
Fig. \ref{fig:stage1principle} shows a simplified model of the nanowire (gray) made of InSb, a material with strong Rashba spin-orbit coupling. Confinement along the wire is created by external voltage-driven gates removed from the picture for clarity. We control voltages applied to these gates to shape the confinement potential, while yet another few gates are used to create electric fields necessary for inducing the Rashba SOI, which is essential for the operation of the device. The initialization scheme consists of two stages.

In the first stage we insert an electron into the quantum wire and trap it inside the confinement potential (Fig. \ref{fig:stage1principle}a). This electron can be of arbitrary spin as the described procedure makes no assumptions on its initial orientation. Next we apply a rising slope of an electric field pulse $E_y$ along the $y$-axis (Fig. \ref{fig:stage1principle}b). This field induces the Rashba SOI and makes the electron wavefunction split into two components of opposite spins which start travelling in opposite directions. As the components travel further they slow down and finally halt for an instant (Fig. \ref{fig:stage1principle}c). At this moment the electric field $E_y$ is maximum and the spin components are somewhat separated but they still overlap. Now, the components start turning back towards the center of the wire due to the repelling influence of the confinement potential. At this very moment, we apply a falling slope of the electric field $E_y$ (Fig. \ref{fig:stage1principle}d). This accelerates the components towards each other and makes them cross unaffected.
This acceleration occurs bacause any change of the electric field $E_y$ affects the components' motion. If the electrons were still moving away from each other the falling slope of $E_y$ would decelerate them but since they have already turned back, the falling slope actually accelerates them further. This peculiar behavior has been described in detail in \cite{ref27}. Now, after the components crossed each other they start slowing down and halt in new positions separated by a distance considerably larger than previously. We need only set a potential barrier between them to lock them in their new positions (Fig. \ref{fig:stage1principle}e). As the components no longer overlap, this indicates full spin separation. This concludes the first stage of initialization. 
\begin{figure}[t]
\includegraphics[clip,trim=0 5em 0 2em,width=0.4\textwidth]{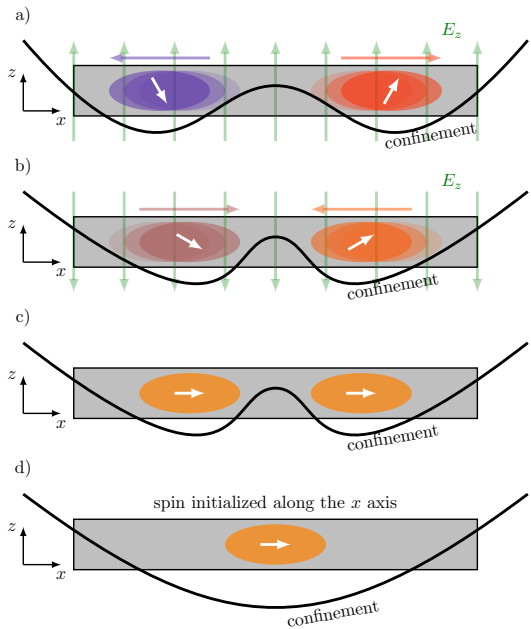}
\caption{\label{fig:stage2principle}Second stage of the spin initialization: (a) By applying the electric field $E_z$ along $z$-axis and modifying positions of the confinement potential minima the spin components' motion along the wire and the spin rotation about $y$-axis are induced; (b) By reverting the positions of minima and reversing the electric field the spins are further rotated and the wavefunction components are brought back to their previous positions; (c) Spins of the both former spin components are now oriented in the same direction effectively ending the initialization procedure; (d) Two wavefunction parts can now be brought back together to create a single wavepacket.}
\end{figure}

Now, the second stage of initialization proceeds, as shown in Fig. \ref{fig:stage2principle}. We start from the point where the previous stage finished. By modifying voltages applied to the gates forming the electron confinement potential, we slightly move the potential minima apart, thus setting both spin components in motion in opposite directions (Fig.~\ref{fig:stage2principle}a). At the same time we apply an electric field $E_z$ along the $z$-axis which induces spin rotations about the $y$-axis. We must not use the term ,,spin components'' anymore as they no longer represent spin up and down with respect to the $z$-axis. From now on, we merely call them wavefunction parts. Because the parts move in opposite directions their spins rotate about the same axis yet by opposite angles. After they travel over some distance they are being brought back to their previous locations by appropriately forming the confinement potential (Fig.~\ref{fig:stage2principle}b). At this point we also have to reverse the $E_z$ direction, because otherwise not only positions but also spins of both parts would revert to their original orientations jeopardizing our efforts. Dividing the spin rotation into two small steps: movement forwards and movement backwards is advantageous as it prevents the wavefunction parts from travelling long distances and allows for simpler control gate layouts. After the spin rotation, the spins of the wavefunction parts are now directed along the $x$-axis, as shown in Fig. \ref{fig:stage2principle}c. Finally, after we turn off the $E_z$ field, both parts can be brought back and merged to form a single wavepacket (Fig. \ref{fig:stage2principle}d). This concludes the second and the last stage of spin initialization.

\section{Simulations}
We have performed time-dependent simulations of nanodevice operation. We use generalized Poisson's equation to solve for the potential $\phi(\mathbf{r},t)$ at every time step in a computational box encompassing the entire nanodevice. The obtained potential is used to calculate the potential energy profile within the quantum wire and along its axis $V(x,t)=-|e|\phi(x,y_0,z_0,t)$ (where $y_0$, $z_0$ are coordinates of the wire), as well as the electric field $\mathbf{E}(\mathbf{r},t)=-\nabla\phi(\mathbf{r},t)$. The time evolution of the electron is obtained by solving the time-dependent Schr\"odinger's equation starting from the electron ground state for the initial potential. A detailed description of the method can be found in \cite{ref26}.

Initially the voltages of top and lateral gates are set to zero $U_\mathrm{top}=U_\mathrm{left}=U_\mathrm{right}=0$. To the remaining seven lower gates we apply $U_{1,\dots,7}=-40\,\mathrm{mV}$, $-10\,\mathrm{mV}$, $-2.5\,\mathrm{mV}$, $0\,\mathrm{mV}$, $-2.5\,\mathrm{mV}$, $-10\,\mathrm{mV}$, $-40\,\mathrm{mV}$. These voltages create a confinement potential energy with nearly parabolic center and high barriers at the borders. 
\begin{figure}[t]
\includegraphics[width=0.4\textwidth]{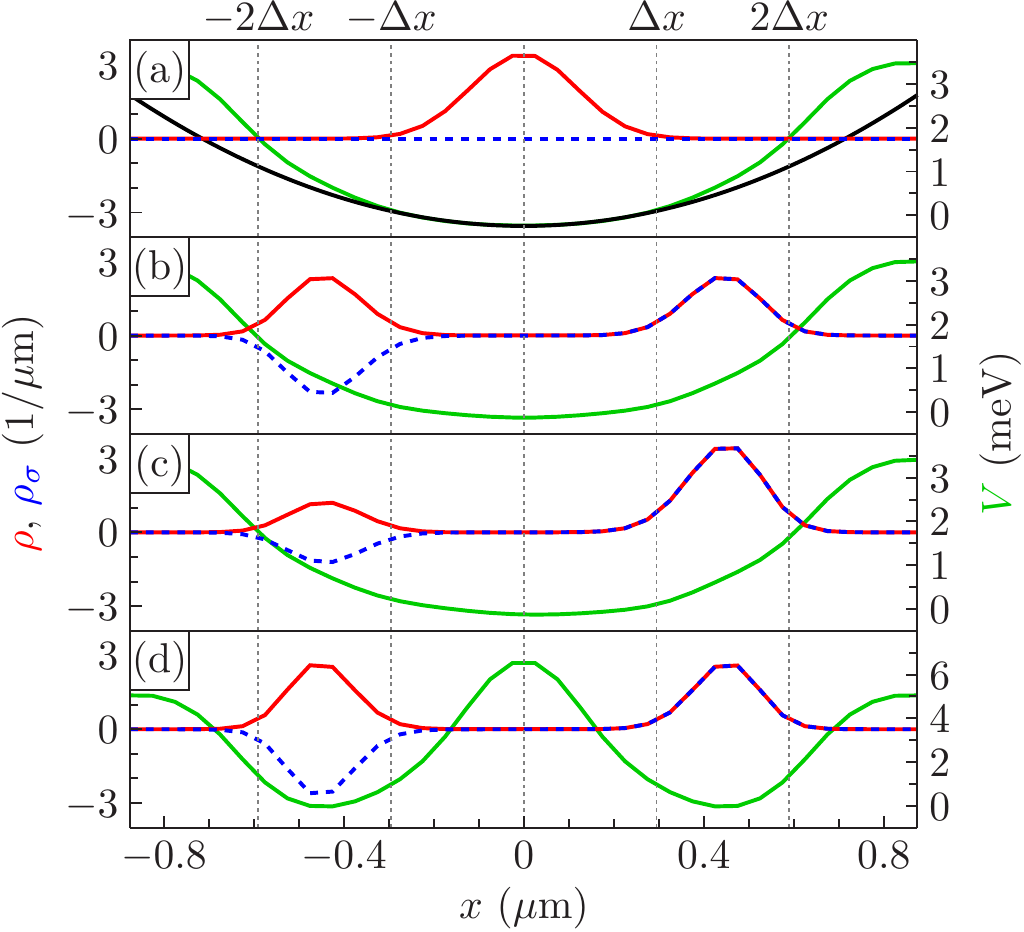}
\caption{\label{fig:wireplot}The electron (red) and spin (blue) densities, together with the electron potential energy (green) along the quantum wire for three selected moments of time: (a) at the beginning, with the black line being a parabola fitted near the potential energy center; (b) after spin separation of an initial state being an equally weighted linear combination of $\mathbf{\Psi}_\uparrow$ and $\mathbf{\Psi}_\downarrow$; (c) just like (b) but with an initial state being an exemplary \emph{non-equally} weighted linear combination of $\mathbf{\Psi}_\uparrow$ and $\mathbf{\Psi}_\downarrow$; (d) after setting up a potential energy barrier between wavepacket parts with opposite spin.} 
\end{figure}
The potential energy profile is shown in Fig. \ref{fig:wireplot}(a) as a green line along with a parabolic fit (black line). The red line marks the charge density (i.e., square of the modulus of the wavefunction $\mathbf{\Psi}^\dagger(x)\mathbf{\Psi}(x)$). We assume that initially the wavefunction corresponds to the electron ground state. 

Because $U_\mathrm{top}=U_\mathrm{left}=U_\mathrm{right}=0$ and the voltages $U_i$ are relatively small, the electric field components $E_y$ and $E_z$ are nearly zero. This effectively causes vanishing of the SOI. Let us now assume that the spin $z$-projection is indefinite and the electron wavefunction is a linear combination of both spin basis states. To the gates $\mathrm{U_{left}}$ and $\mathrm{U_{right}}$ we apply a single voltage pulse lasting half a period, given by the formulae $U_\mathrm{left}(t)=-U_\mathrm{sep}\sin(\omega_\mathrm{sep}t)$ and $U_\mathrm{right}(t)=U_\mathrm{sep}\sin(\omega_\mathrm{sep}t)$ where $t\in[0,\pi/\omega_\mathrm{sep}]$, the amplitude $U_\mathrm{sep}=1.3\,\mathrm{V}$ and $\hbar\omega_\mathrm{sep}=0.15\,\mathrm{meV}$. This pulse generates an electric field parallel to the $y$-axis, and equivalently the Rashba SOI, which causes spatial separation of the wavefunction into two parts with opposite spin directions.
\begin{figure}[h]
\includegraphics[width=0.4\textwidth]{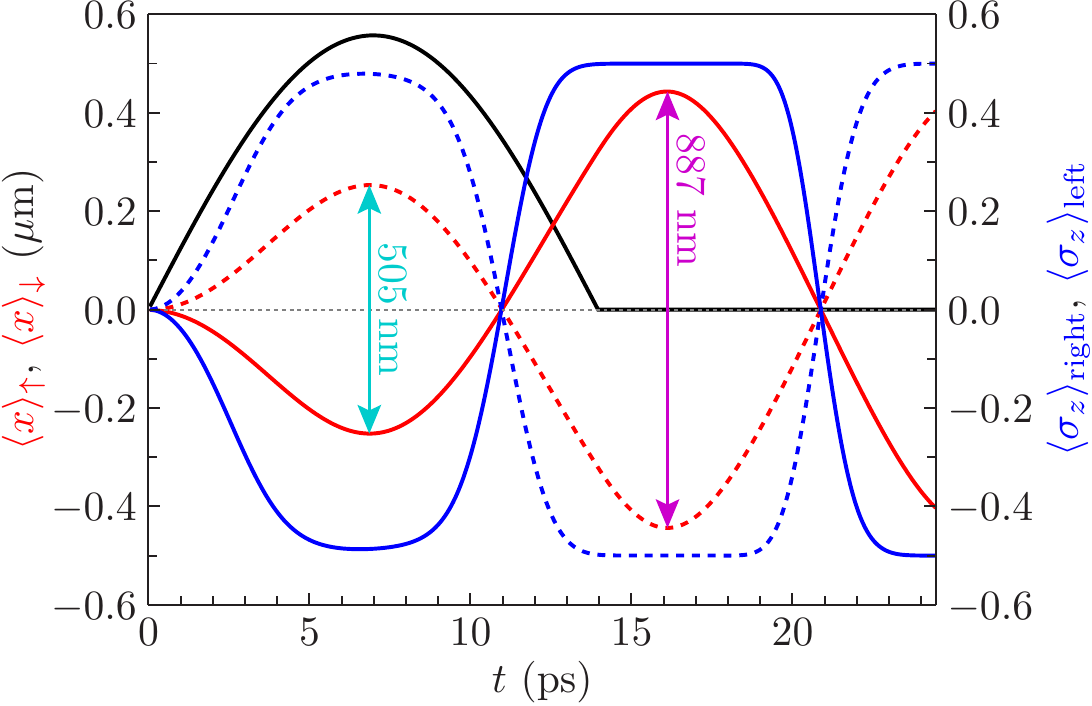}
\caption{\label{fig:separation}Time courses in the spin separation stage. The black curve shows the voltage pulse applied to gate $\mathrm{U_{left}}$. The red curves (solid and dashed) show the expectation values of position of both spin parts of the wavefunction ($\langle x\rangle_\uparrow$ and $\langle x\rangle_\downarrow$), while the blue curves show the expectation values of the Pauli $z$-matrix calculated in the right ($\langle\sigma_z\rangle_\mathrm{right}$) and left ($\langle\sigma_z\rangle_\mathrm{left}$) halves of the nanodevice.}
\end{figure}

Fig.~\ref{fig:separation} shows results of the electron state time evolution. The black curve shows a time course of $U_\mathrm{left}(t)$, the solid red curve the expectation value of position calculated for the upper spinor part as $ \langle x\rangle_\uparrow=\langle\psi_\uparrow|\hat{x}|\psi_\uparrow\rangle/\langle\psi_\uparrow|\psi_\uparrow\rangle$ and the dashed red curve $\langle x\rangle_\downarrow$ calculated in a similar way using the lower spinor part $\psi_\downarrow$. Initially, at $t=0$, both $x_\uparrow$ and $x_\downarrow$ are identical and correspond to the middle of the wire where the potential energy minimum is located. An increasing voltage between $\mathrm{U_{left}}$ and $\mathrm{U_{right}}$ induces separation of both wavepacket parts. At $t=7\,\mathrm{ps}$, when the voltage pulse reaches its maximum, the wavepacket parts cease to separate any more and start moving backwards. At this moment, the spacing between both parts calculated from Eq. (\ref{eq:displacement}) equals $2\Delta x = 590\,\mathrm{nm}$. This agrees only approximately with the more accurate value $505\,\mathrm{nm}$ obtained from the simulation shown in Fig. \ref{fig:separation} (marked with a cyan arrow). This is so, because Poisson's equation solved in the simulations additionally takes into account the interaction of the electron with charge induced on local gates. From now on, the spin orbit coupling decreases but still accelerates both wavepacket parts, since they bounced off of the potential energy barriers and move in the opposite directions\cite{ref27}. As a result, the wavepacket energy continues to grow until the voltage pulse returns to zero. For an exactly parabolic potential energy, the spacing between wavepacket parts would be twice as large (and equal to $4\Delta x$), but the actual potential energy does not satisfy this condition and changes its shape during the wavepacket separation.

Note that the entire process is not resonant and $\omega_\mathrm{sep}$ does not have to be equal to $\omega$ from Eq. \ref{eq:potential}. Because the assumed value of $\omega_\mathrm{sep}$ is approximately $20\%$ lower than $\omega$, in Fig. \ref{fig:separation} the opposite spin wavepackets return to the initial position sooner (at $t=11\,\mathrm{ps})$ than the voltages fall down to zero (at $t=14\,\mathrm{ps}$).

The solid blue curve in Fig.~\ref{fig:separation} depicts the expectation value of the Pauli $z$-matrix, calculated in the right half of the quantum wire as (note the limits):
\begin{equation}\label{eq:rightspin}
\langle\sigma_z\rangle_\mathrm{right}=\int_0^{L/2}\mathbf{\Psi^\dagger}(x,t)\sigma_z\mathbf{\Psi}(x,t)dx.
\end{equation}
The value of $\langle\sigma_z\rangle_\mathrm{left}$, calculated in a similar way, is shown in Fig. \ref{fig:separation} as a dashed blue line. In the presented simulation the initial electron state was an equally weighted linear combination of spin states (i.e., $c_\uparrow=c_\downarrow=1/\sqrt{2}$), thus at $t=15\,\mathrm{ps}$ $\langle\sigma_z\rangle_\mathrm{right}=0.5$ and $\langle\sigma_z\rangle_\mathrm{left}=-0.5$. This indicates a full spatial separation of spin parts, as shown in Fig. \ref{fig:wireplot}(b). If the initial linear combination of spin states was not equally weighted, the final values of $|\langle\sigma_z\rangle_\mathrm{right}|$ and $|\langle\sigma_z\rangle_\mathrm{left}|$ would not be equal. This situation is shown in Fig.~\ref{fig:wireplot}(c).
In the most extreme case, when spin is oriented upwards, i.e. $c_\uparrow=1$ (or downwards, $c_\downarrow=1$), the electron will occupy the right (or left) half of the quantum wire with probability 1. Let us note that this stage of operation can also be used for spin readout. This operation can be performed in $T_{\mathrm{READOUT}}=15\,\mathrm{ps}$.

Fig.~\ref{fig:wireplot}(b) shows the electron density along the wire (red curve) calculated as $\rho(x,t)=\mathbf{\Psi^\dagger}(x,t)\mathbf{\Psi}(x,t)=|\psi_\uparrow(x,t)|^2+|\psi_\downarrow(x,t)|^2$ and spin density (blue curve) as $\rho_\sigma(x,t)=\mathbf{\Psi^\dagger}(x,t)\sigma_z\mathbf{\Psi}(x,t)=|\psi_\uparrow(x,t)|^2-|\psi_\downarrow(x,t)|^2$. According to these definitions in the region where spin is directed upwards the curves overlap, which occurs in the right side of the nanodevice, while for spin directed downwards they have opposite signs, which occurs in the left. At the moment when the distance between wavepackets is maximal, we change gate voltages appropriately, creating a potential barrier between them and confining them inside two separate potential valleys. The barrier has to be sufficiently high, and the minima deep, so as to allow independent operations on spin in both valleys. To achieve this, at $t_1=16\,\mathrm{ps}$ we change the gate voltages rapidly to $U_{1,\dots,7}(t_1)=-60\,\mathrm{mV}$, $10\,\mathrm{mV}$, $-10\,\mathrm{mV}$, $-100\,\mathrm{mV}$, $-10\,\mathrm{mV}$, $10\,\mathrm{mV}$, $-60\,\mathrm{mV}$. The obtained potential energy profile as well as electron and spin densities are plotted in Fig. \ref{fig:wireplot}(d). The potential energy has two minima with a barrier between them, separating the wavefunction spatially into two parts of opposite spin directions.

Now we proceed to the second stage of operation which boils down to a spin rotation about the $y$-axis. If we turn spin in the right valley clockwise by $\pi/2$ and in the left -- counterclockwise by the same angle, spins of both parts become parallel to each other and directed along the $x$-axis. From the form of Hamiltonian (Eq. \ref{eq:hamiltonian}), it follows that the motion along the $x$-axis induces spin rotation about the $y$-axis if an electric field $E_z$ (along the $z$-axis) is present. We generate it by applying a voltage to top gate. Now spin rotation is achieved by setting the electron in an oscillatory motion along the $x$-direction.

To achieve this, from the time $t_1$ onwards, voltages applied to gates $\mathrm{U_2}$ and $\mathrm{U_3}$ are modified according to the formulae: $U_2(t)=U_2(t_1)-U_\mathrm{osc}\left(1-\cos\left(\omega_\mathrm{rot}(t-t_1)\right)\right)$ and $U_3(t)=U_3(t_1)+U_\mathrm{osc}\left(1-\cos\left(\omega_\mathrm{rot}(t-t_1)\right)\right)$. Because we want to rotate spin of both wavepacket parts in opposite directions, to gates $\mathrm{U_5}$ and $\mathrm{U_6}$ we must apply voltages stimulating motion in the opposite direction: $U_5(t)=U_3(t)$ and $U_6(t)=U_2(t)$. At the same time $t=t_1$ we turn on the Rashba SOI, then reverse its sign when the wavepacket parts stop and start moving backwards. This is achieved by applying a voltage, phase-shifted by $\pi/2$, to top gate: $U_\mathrm{top}(t)=-U_\mathrm{rot}\sin(\omega_\mathrm{rot}(t-t_1))$ \cite{ref28}. In this simulation stage, the pulse duration equals one full period of sine. Using a pulse lasting half a period did not produce satisfactory fidelity of spin initialization. The value $\omega_\mathrm{rot}$ does not have to be carefully selected, but it should not be too small, as it affects the duration of operation and not too large, because with its increase fidelity drops. In the simulations we assumed a value of $\hbar\omega_\mathrm{rot}=0.11\,\mathrm{meV}$. The amplitude $U_\mathrm{osc}$ of voltages stimulating wavepacket oscillations does not have to be precisely chosen and we assumed $U_\mathrm{osc}=100\,\mathrm{mV}$. Also, the exact shape of pulses is not critical and deviations from the presented sine-like shapes is acceptable. Only the $U_\mathrm{rot}$ amplitude must be tuned to the pulse duration. Performed simulations indicate, that to maintain fidelity at the level of 99\% or greater, $U_\mathrm{rot}$ should be selected with accuracy better than $\pm 40\,\mathrm{mV}$.

% Only one parameter must be adjusted, namely the amplitude $U_\mathrm{rot}=930\,\mathrm{mV}$ (tuned for the chosen value of $\hbar\omega_\mathrm{rot}$). Performed simulations indicate, that to maintain fidelity at a level of $99\%$ or greater, $U_\mathrm{rot}$ should be selected with accuracy better than $\pm 40\,\mathrm{mV}$.

\begin{figure}[h]
\includegraphics[width=0.4\textwidth]{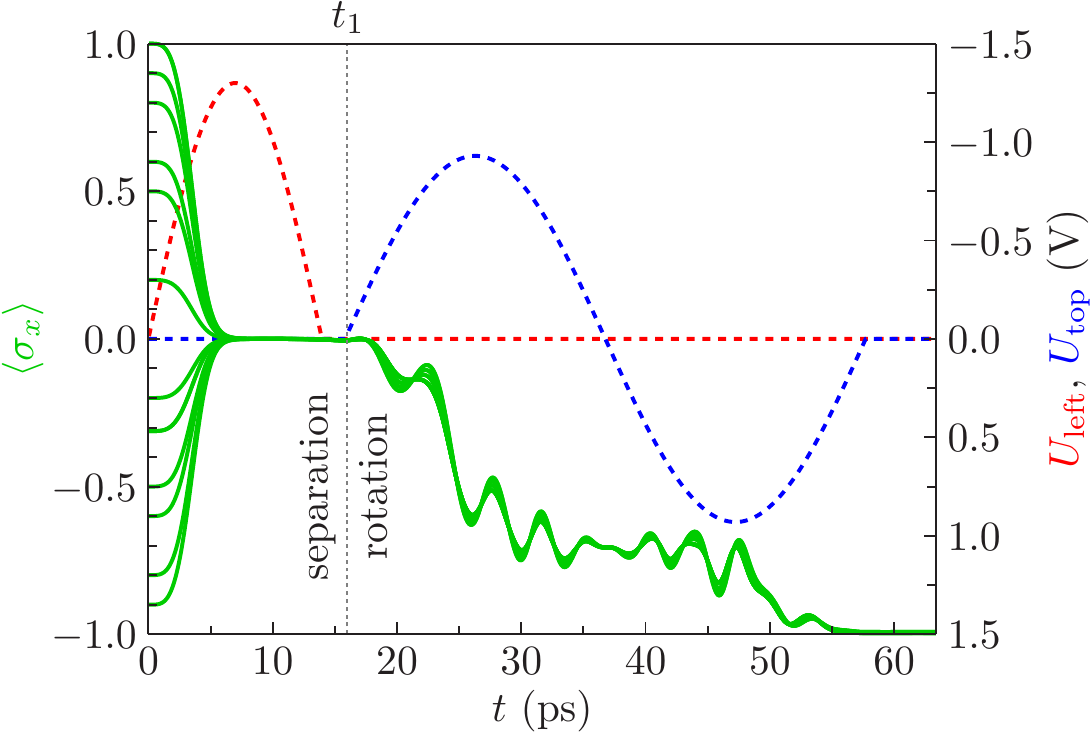}
\caption{\label{fig:FIG4}Time courses of expectation values of the Pauli $x$-matrix, namely $\langle\sigma_x\rangle$, for various initial spin orientations (green). The red dashed curve shows the separating voltage pulse, while the blue dashed curve shows the pulse, responsible for SOI, used for spin rotations in both parts of the device.}
\end{figure}

Fig. \ref{fig:FIG4} shows time evolutions of expectation values of the Pauli $x$-matrix $\langle\sigma_x\rangle$ for several initial spin orientations. The courses differ considerably only in the first stage of nanodevice operation, lasting about $7\,\mathrm{ps}$. At $t=t_1=16\,\mathrm{ps}$, the wavepacket is split into two parts, one of which has spin parallel to the $z$-axis, while the second one spin antiparallel. At this moment $\langle\sigma_x\rangle$ vanishes. In the second stage of simulation, in which spin parts are rotated, the courses overlap regardless of the initial spin orientations and all reach a value close to unity at the same time $t=60\,\mathrm{ps}$. The final fidelity of spin initialization is of the order of $99.3\%$.

After the entire procedure, spin becomes oriented along the $x$-direction. However, if we want to further change its orientation, this can be done using another voltage pulse. Voltages applied to the lateral gates can generate the SOI, which induces spin rotation about the $z$-axis. On the other hand a voltage applied to the top gate allows for rotations about the $y$-axis.

The second pulse, visible in Fig. \ref{fig:FIG4}, lasting about $45\,\mathrm{ps}$ resulted in a spin rotation by $90^\circ$. Such operation is performed by the Hadamard gate, and in our case it takes $T_\mathrm{HADAMARD}=45\,\mathrm{ps}$. The NOT gate requires a rotation by $180^\circ$, thus it requires $T_\mathrm{NOT}=90\,\mathrm{ps}$. During the second stage of the presented initialization scheme we rotate spin in the right potential valley clockwise and in the left -- counterclockwise. If spin was rotated by $180^\circ$ only in the left valley, leaving the right one unchanged, what can be achieved by modulating voltages $\mathrm{U_2}$ and $\mathrm{U_3}$ and fixing voltages $\mathrm{U_5}$ and $\mathrm{U_6}$, spin of the electron would be reversed if it occupied the left valley or remained untouched if it occupied the right one. This outcome is equivalent to the controled negation (CNOT) two-qubit gate if we assume that the first qubit is a spin qubit while the second is a charge qubit defined as presence of an electron in the left or right potential valley. The operation time of this gate is $T_\mathrm{CNOT}=90\,\mathrm{ps}$.

\section{Summary}
We proposed a nanodevice designed to set spin of a single electron in a desired direction. After two pulses of voltages, lasting less than $60\,\mathrm{ps}$ in total, spin is set in parallel to the $x$-axis. The outcome does not depend on initial spin orientations and is obtained without using any external fields, microwaves or photons. The goal is achieved all electrically with voltages applied to the local gates.
The proposed nanodevice can also be used to perform other necessary quantum operations: readout, Hadamard gate, NOT gate, CNOT gate in times: $15\,\mathrm{ps}$, $45\,\mathrm{ps}$, $90\,\mathrm{ps}$ and $90\,\mathrm{ps}$, respectively. These estimated operation times, compared to the coherence time of about $34\,\mathrm{ns}$ look very promising.

In the performed simulations the nanodevice is modeled upon other similar nanostructures described in experimental papers, which assures its experimental feasibility. We used real material parameters like the InSb electron effective mass or distinct dielectric constants of the nanowire and the surrounding insulator. The assumed electric pulse durations are short yet experimentally achievable, while their amplitudes are low enough to ensure adiabaticity of the entire process. The estimated operation times are extracted directly from the time courses in Fig. \ref{fig:FIG4}. They are achievable in practice but shortening them any further might prove difficult.

\section{Acknowledgements}
This work has been supported by National Science Centre (NSC), Poland, under UMO-2014/13/B/ST3/04526. JP has been supported by National Science Centre, under Grant No. 2016/20/S/ST3/00141. This work was also partially supported by the Faculty of Physics and Applied Computer Science AGH UST dean grants No. 15.11.220.717/24 and 15.11.220.717/30 for PhD students and young researchers within subsidy of Ministry of Science and Higher Education.


\end{document}